\pdfoutput=1
\documentclass[openacc]{rsproca_new}

\usepackage{microtype}

\def\titlehead{Research}

\jname{rspa}
\Journal{Proc R Soc A\ }

\begin{document}

\title{Can comets deliver prebiotic molecules to rocky exoplanets?}

\author{
R. J. Anslow$^{1}$, A. Bonsor$^{1}$ and P. B. Rimmer$^{2}$}

\address{$^{1}$Institute of Astronomy, University of Cambridge, Madingley Road, Cambridge CB3 0HA, UK\\
$^{2}$Astrophysics Group, Cavendish Laboratory, University of Cambridge, JJ Thomson Ave, Cambridge, CB3 0HE, UK}

\subject{Astrobiology, Extrasolar planets}

\keywords{Origins of life, prebiotic chemistry, planetary dynamics, hydrogen cyanide, habitable planets, impacts}

\corres{R. J. Anslow\\
\email{rja92@ast.cam.ac.uk}}

\begin{abstract}
In this work we consider the potential of cometary impacts to deliver complex organic molecules and the prebiotic building blocks required for life to rocky exoplanets. Numerical experiments have demonstrated that for these molecules to survive,  impacts at very low velocities are required. This work shows that for comets scattered from beyond the snow-line into the habitable zone, the minimum impact velocity is always lower for planets orbiting Solar-type stars than M-dwarfs. Using both an analytical model and numerical N-body simulations, we show that the lowest velocity impacts occur onto planets in tightly-packed planetary systems around high-mass (i.e.\ Solar-mass) stars, enabling the intact delivery of complex organic molecules. Impacts onto planets around low-mass stars are found to be very sensitive to the planetary architecture, with the survival of complex prebiotic molecules potentially impossible in loosely-packed systems. Rocky planets around M-dwarfs also suffer significantly more high velocity impacts, potentially posing unique challenges for life on these planets. In the scenario that cometary delivery is important for the origins of life, this study predicts the presence of biosignatures will be correlated with i) decreasing planetary mass (i.e.\ escape velocity), ii) increasing stellar-mass, and iii) decreasing planetary separation (i.e.\ exoplanets in tightly-packed systems).
\end{abstract}


\maketitle


\section{Introduction}
\label{sec:introduction}
The initial emergence of life on Earth will require some initial inventory of prebiotic molecules, for which there are two leading supply mechanisms, endogenous synthesis on the early-Earth and exogenous delivery. There are a number of synthesis pathways, such as via lightning discharge \cite{ChameidesWalker1981, Barth2023}, atmospheric proton irradiation \cite{Kobayashi2023}, atmospheric photochemical networks \cite{Zahnle1986, Tian2011}, and the shock-synthesis of carbonaceous and nitrogenous molecules during high velocity impacts \cite{Ferus2017, Parkos2018}. The efficacy of these pathways, however, is dependent on the atmospheric oxidation state of the early-Earth, decreasing significantly in more oxidised atmospheres \cite{RimmerRugheimer2019}.

The exogenous delivery of prebiotic molecules is an attractive, atmosphere-independent supply mechanism encompassing the delivery of extraterrestrial objects including asteroids, comets and interplanetary dust particles \cite{Oro1961, Anders1989, Chyba1990, ChybaSagan1992}. Return samples from the Ryugu asteroid have revealed a diverse inventory of prebiotic molecules \cite{Oba2023, Naraoka2023}, with the discovery of intact amino acids in meteorite samples \cite{Kvenvolden1970, CroninMoore1971} highlighting that at least some prebiotic molecules are able to survive atmospheric entry. The survival of prebiotic molecules in meteorites may not even be necessary, given that carbonaceous chondrites are able to catalyze a range of prebiotic molecules in aqueous conditions, and thereby fertilise the early-Earth \cite{Rotelli2016, Trigo-Rodriguez2019}.

Comets have been suggested as potentially important sources of prebiotic delivery (e.g. \cite{PierazzoChyba1999}) as they are known to contain large amounts of the prebiotic feedstock molecule hydrogen cyanide (HCN) \cite{MummaCharnley2011}, as well as simple amino acids \cite{Elsila2009}. The discovery of a rich diversity of CHN- and CHS-bearing molecules on the comet 67P \cite{Altwegg2017} has further supported the potential importance of cometary impacts in delivering part of the early-Earth's organic inventory. Despite the relatively small number of cometary impacts on the early-Earth (in comparison to asteroids and left-over planetesimals) \cite{Sinclair2020, Nesvorny2023}, it has been estimated that comets delivered 2 orders of magnitude more organic material than meteorites \cite{ChybaSagan1992}. This is a consequence of the extremely high carbon content of comets ($\sim 10$\% \cite{Anders1989, MummaCharnley2011}) in comparison to both C-and S-type asteroids (2\% \cite{Sephton2002} and 0.2\% \cite{MooreLewis1967} respectively).

There has, however, been longstanding debate surrounding the plausibility and efficiency of cometary delivery given the pyrolysis, or thermal decomposition, of constituent organics during impacts (e.g. \cite{PierazzoChyba1999}), and the subsequent dilution of surviving molecules in the atmosphere and oceans. The `warm comet pond' \cite{Clark1988} has been proposed as a specific origins scenario, able to alleviate both of these concerns and support the initial emergence of life. The scenario requires the `soft-landing' of a cometary nucleus, which excavates the impact point and forms a dirty pond from the cometary components. Climatic variations are thought to cause the episodic drying of these ponds, promoting the rapid polymerisation of constituent prebiotic molecules. It is thought this wet-dry cycling will effectively drive the required biogeochemical reactions crucial for RNA production on the early-Earth, and therefore play an important role in the initial emergence of life (e.g. \cite{Pearce2017, Campbell2019}). Relatively high concentrations of prebiotic molecules are required for there to be sufficient polymerisation, and so this scenario still requires low-velocity impacts. Specific prebiotic molecules are more (or less) susceptible to thermal decomposition by virtue of their molecular structure, and so the inventory of molecules that can be effectively delivered to a planet is very sensitive to impact velocity. 

HCN is an example of a prebiotic molecule that is particularly suited to this warm comet pond scenario, as its strong carbon-nitrogen triple bond confers much greater durability to the temperatures experienced during impacts. Given also the abundance of HCN in comets \cite{MummaCharnley2011}, cometary delivery onto the early-Earth is thought to have been able to successfully deliver high concentrations of HCN to local environments, which can survive over 0.1-1\,Myr timescales \cite{ToddOberg2020}. This has significant biological implications, since HCN is a key feedstock molecule in multiple prebiotic syntheses, as a precursor to amino acids through Strecker synthesis \cite{Strecker1854}, invoked in the Miller-Urey experiments \cite{Miller1953}. Further studies have demonstrated the role of HCN in the synthesis of lipids, sugars, nucleobases and nucleotides \cite{Ferus2015, Patel2015, Xu2018}, four key building blocks of life, as summarised in \cite{Sutherland2016}. The strict velocity constraints required in the comet pond scenario naturally prompt the question of whether the Earth is somewhat special in this regard.

Exoplanets are observed across a wide range of environments, with M-dwarf stars the most numerous spectral-class in our galaxy, comprising $\sim$70\% of all stars in the Solar-neighbourhood \cite{Bochanski2010}. Many of these stars host large outer reservoirs of planetesimals in belts analogous to the Solar System asteroid and Kuiper belts, which generate infrared excesses through the collisional production of dust \cite{Wyatt2008, Eiroa2013}. These belts are inferred to have masses $\sim 10^2 M_\mathrm{Earth}$, roughly 6 orders of magnitude larger than the Kuiper belt \cite{KrivovWyatt2020}. The detection of carbon monoxide gas in several of these systems indicates the presence of volatile-rich cometary bodies, suggesting a similarity to Solar System comets \cite{ZuckermanSong2012, Marino2016}. There are relatively few known debris discs around M-dwarfs, but systems such as the nearby M-dwarf AU Microscopii, with its two warm-Neptunes \cite{Plavchan2020, Martioli2021} orbiting interior to a cold, dusty debris belt \cite{Kalas2004}, provide key examples that architectures similar to that proposed here do exist.

Thus, in this study we hypothesize that comets host comparable prebiotic molecular diversity in other planetary systems, and ask the question "Which planets, and planetary systems are the most susceptible to successful cometary delivery?". The detection of habitable planets around F- and G-type stars remains a formidable observational challenge, and there exist unique challenges facing the habitability of planets around both M-dwarfs and F-type stars \cite{Sato2014, Shields2016}. Nonetheless, by addressing this question, we can begin to test the hypothesis that cometary delivery plays an important role in origins scenarios against data from the next generation of great observatories.

The paper starts in Section~\ref{sec:planetary_architecture} by introducing an idealised planetary architecture based on observations of tightly-packed multiple planet systems. We present in Section~\ref{sec:theory_methods} an analytical method to determine how slowly comets are potentially able to impact onto rocky habitable planets. Section~\ref{sec:num_sims} describes the numerical simulations that will be used to confirm this result and investigate how frequently low velocity impacts can occur. In Section~\ref{sec:results} we present the results of these simulations in the context of our analytical predictions. In Section~\ref{sec:discussion} we discuss the results of our analytical model and simulations, and present predictions for future correlations between biosignatures and exoplanet populations. Finally, Section~\ref{sec:conclusions} summarises our conclusions.

\section{Methods}
\label{sec:methods}
\begin{figure}[!t]
    \centering
    \includegraphics[width=0.9\textwidth]{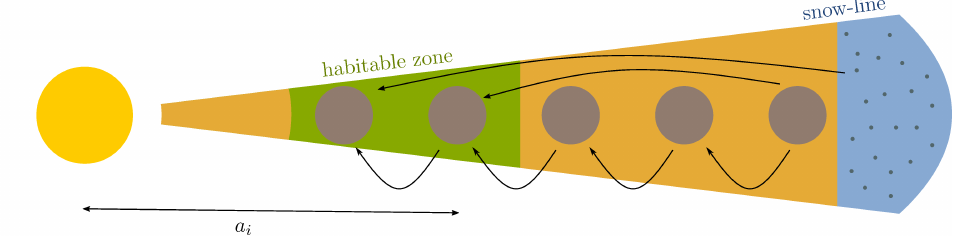}
    \caption{Schematic diagram of the idealised planetary system considered in this work with equally spaced planets (brown circles, semi-major axis $a_i$) scattering comets (small dark blue circles) from the snow-line. The blue region represents the volatile-rich region of the disc where comets occur, and the green region represents the habitable zone. Low velocity cometary impacts onto habitable planets will follow the lower arrows, which sketch the dynamically cold scattering between adjacent planets. The dynamically hot scattering as shown by the upper arrows, will result in high velocity impacts.}
    \label{fig:planet_chain_schematic}
\end{figure}

The cometary delivery of prebiotically interesting molecules requires very low impact velocities, to minimise the amount of thermal degradation and allow a substantial quantity to survive the impact. To understand the importance of cometary delivery in an exoplanetary context, we study the scattering of volatile-rich bodies from beyond the snow-line into the habitable zone. We consider an idealised planetary system with equally spaced and equal-mass planets, as sketched in figure~\ref{fig:planet_chain_schematic}, which allows us to make analytical predictions for the minimum impact velocity onto the inner habitable planet. We then use N-body simulations to study the overall changes to the velocity distribution and test our analytical predictions.

\subsection{Idealised planetary architecture}
\label{sec:planetary_architecture}
We consider a chain of co-planar planets, each with mass $M_\mathrm{Earth}$, on circular orbits around a central star (mass $M_\ast \in$ [0.1, 2]\,$M_\mathrm{Sun}$). The semi-major axis of the inner planet is fixed in the centre of the star's habitable zone, which is calculated using the stellar mass-luminosity relationship and habitable zone prescription from \cite{Scalo2007, Kopparapu2013} respectively. The semi-major axes of the other planets are chosen such that the planets are equally spaced in units of their mutual Hill radius, 
\begin{align}
    \label{eq:delta_a_equal_spacing}
    a_{i+1} &= a_i + \Delta R_{H_{i,i+1}}, \\
    \label{eq:mutual_hill_radius}
    R_{H_{i, i+1}} &= \left(\frac{2M_\mathrm{Earth}}{3M_\ast}\right)^{1/3} \left(\frac{a_i + a_{i+1}}{2}\right) \equiv X \left(a_i + a_{i+1}\right).
\end{align}
We are interested in the scattering of volatile-rich material onto the inner planet, and so use the snow-line to delineate the boundary between the populations of rocky and volatile-rich bodies in the disc. The number of planets is chosen to span the gap between the habitable zone and snow-line. The location of the snow-line is calculated under the assumption of an optically thick, steady-state protoplanetary disc as outlined in \cite{Alcala2014, Min2011}.

This choice of architecture is inspired by the prevalence of so-called 'peas in a pod' systems \cite{Weiss2022}. This describes the observed phenomenon that individual exoplanet systems have much smaller dispersion in mass, radius, and orbital period in comparison to the system-to-system variation of the exoplanet population as a whole \cite{Weiss2018, Millholland2017, Wang2017}. The observed correlation between the size of the planets and their orbital period ratios implies that gravitational interactions between neighboring planets are responsible for this effect, for which the mutual Hill radius is the natural length scale. 

Whilst this chosen architecture is clearly idealised, numerical simulations have demonstrated that pebble accretion may naturally form these tight multi-planet systems, in addition to outer giant planets \cite{Lambrechts2019, Bitsch2019}. These giant planets may be an important source of comets into the inner planetary system, given their efficiency at scattering comets onto habitable zone-crossing orbits \cite{DencsRegaly2019}. In this work, however, we do not specify the outer planetary architecture in an attempt to keep the model as general as possible.

In the \textit{Kepler} systems, the interplanetary separations typically range between 10-30 mutual Hill radii ($\Delta$) \cite{Weiss2018, PuWu2015}. Below 10\,$\Delta$ the stability dramatically decreases, due to the high degree of dynamical complexity present in these tightly-packed systems \cite{Chambers1996, Obertas2017}. For separations below a critical value $\Delta_c$ \cite{Gladman1993, Deck2013}, these systems are stable for less than 10$^2$ conjunctions. We therefore consider planetary spacings in the range 10 to 80 mutual Hill radii, which spans both the Kepler systems and terrestrial Solar System planets. Numerical simulations have demonstrated that tightly-packed systems with equal and low mass planets are particularly efficient at scattering comets into the innermost planets \cite{Marino2018}. These results ensure efficient cometary scattering into the habitable zone, and are used to guide the range of spacings we consider in section~\ref{sec:num_sims}.

\subsection{Analytical predictions}
\label{sec:theory_methods}
Simulations have demonstrated that the scattering of particles by multiple planets can be well-approximated by a series of three-body interactions \cite{LevisonDuncan1997}, between which the orbits of the particles conserve their Tisserand parameter \cite{Tisserand1889, MurrayDermott1999}. We therefore separate the dynamics of particles into two distinct regions - inside and outside of the scattering planet's Hill sphere. Inside the Hill sphere we assume the particle feels only the attraction of the central planet, so is pulled onto a hyperbolic orbit with the planet at the focus. With $v_\infty$ the relative velocity to the planet "at infinity", the impact velocity onto the planet is given by $v_\mathrm{imp}^2 = v_\mathrm{esc}^2 + v_\infty^2$. The relative velocity at infinity is determined by the particle's orbital elements outside of the Hill sphere, and is given by \cite{Opik51}
\begin{equation}
    v_\infty^2 = v_\mathrm{pl}^2  \left(3 - \frac{a_\mathrm{pl}}{a} - 2\cos{i}\sqrt{\frac{a(1-e^2)}{a_\mathrm{pl}}}\right) = v_\mathrm{pl}^2 \left(3 - \mathcal{T} \right).
\end{equation}
There are therefore three key factors to consider when determining the minimum impact velocity; the planet's escape velocity, the planet's orbital (Keplerian) velocity, and the particle's orbit relative to the planet, as characterised by the Tisserand parameter, $\mathcal{T}$. The maximum Tisserand parameter of the particle is constrained by the planetary architecture, corresponding to an orbit with apocentre $Q = a_1$, and pericentre $q = a_0$. We are able to analytically calculate both the maximum Tisserand parameter, and minimum impact velocity for our idealised planetary architecture, 
\begin{align}
    \label{eq:maximum_tiss_param}
    \mathcal{T}_\mathrm{max} &= \left(1 - \Delta X\right) + 2\sqrt{1 + \Delta X}, \\
    \label{eq:minimum_imp_vel}
    v_{\mathrm{imp}, \mathrm{min}}^2 &= v_\mathrm{esc}^2 + \left(\frac{GM_\ast}{d_\mathrm{HZ}}\right) \left(2 + \Delta X - 2\sqrt{1 + \Delta X}\right),
\end{align}
where $d_\mathrm{HZ}=d_\mathrm{HZ}\left(M_\ast\right)$ is the radial location of the habitable zone.

\subsection{Numerical simulations}
\label{sec:num_sims}
The overall velocity distribution onto a planet will be determined by the inwards flux of particles (on a range of orbits), and the accretion cross section of the planet\footnote{The accretion cross section of a planet is a function of the planetary and stellar properties, and the particle's orbit. It is given by $\sigma_\mathrm{acc} \propto R_\mathrm{pl}^2 \left(1 + v_\mathrm{esc}^2/v_\infty^2\right)$, which we note is a function of the particle's Tisserand parameter.}. The location of planets in $(a_\mathrm{pl}, M_\mathrm{pl})$ space determines the dynamics of the scattered particles, with the equality of the escape velocity and Keplerian velocity roughly constraining the efficiency of particle ejection from the system \cite{Wyatt2017}. The details of the planetary architecture is therefore able to dramatically alter the inwards flux and relative velocities of the scattered particles in the system. Whilst the Tisserand parameter is conserved during interactions with a single planet, it is not conserved in the scenario that a particle has multiple interactions and is exchanged between two scattering planets \cite{BonsorWyatt2012}, making it challenging to make analytical predictions about the overall velocity distribution. We therefore turn to N-body simulations, which are well-suited to studying the effects of both the planetary architecture and stellar-mass on the overall velocity distribution.

N-body simulations were carried out using the MERCURIUS hybrid integrator \cite{Rein2019} from the open source code REBOUND \cite{ReinLiu2012}. This integrator allows us to speed up the long-term integrations during distant interactions, without losing precision during close encounters. The scheme uses the 15th order adaptive integrator IAS15 \cite{ReinSpiegel2015} during close encounters, and otherwise uses the symplectic Wisdom-Holman integrator WHFast \cite{WisdomHolman1991, ReinTamayo2015}. All planets are chosen to have the same mass and radius as the Earth, and we adopt a timestep of $P_0 / 50$, where $P_0$ represents the period of the innermost planet. The inner planet is fixed in the star's habitable zone, and the semi-major axes of the scattering planets are given by
\begin{equation}
    a_n = a_0 \left(\frac{1 + \Delta X}{1 - \Delta X}\right)^n.
\end{equation}
The number of planets, $n$, is chosen to fully span the gap between the habitable zone and snow-line (i.e. $a_n > d_\mathrm{SL}$). The conservation of the Tisserand parameter determines the minimum pericentre each test-particle can be scattered to \cite{BonsorWyatt2012}, and so we use the requirement for planet-crossing orbits to generate the test-particles' Tisserand parameter distribution. We inject test-particles into the chaotic zone of the outer planet \cite{WisdomHolman1991}, and record the impact velocities when the particles collide with the physical radius of each planet in the simulation.

\section{Results}
\label{sec:results}
In this section we first discuss our analytical predictions for the minimum impact velocities onto the habitable planet in our idealised planetary system. We then present the results of the N-body simulations, and discuss how the overall velocity distribution is affected by the stellar-mass and planetary architecture. Our analytical predictions are then compared with the N-body simulations to verify their accuracy. 

\subsection{Analytical results}
\begin{figure}[t!]
    \centering
    \includegraphics[width=\textwidth]{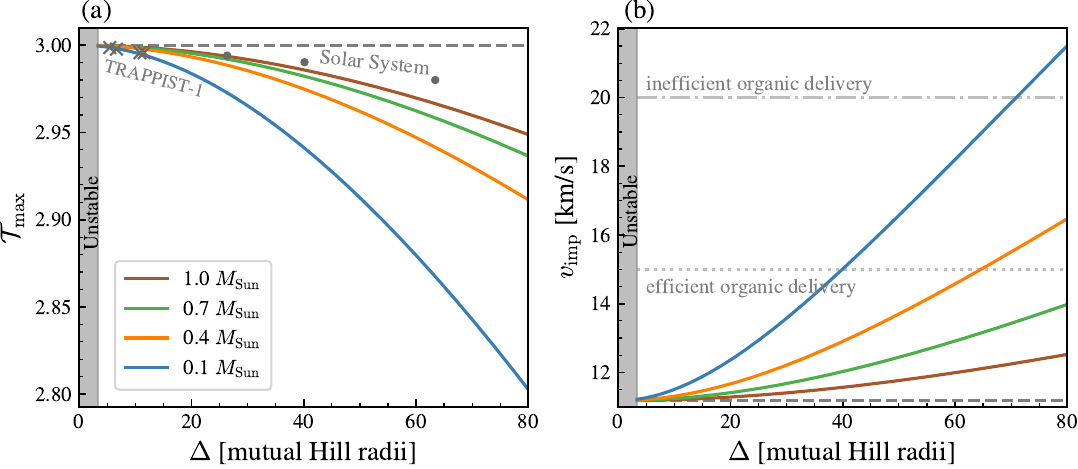}
    \caption{\textbf{Left panel.} Analytical predictions for the maximum Tisserand parameter of comets onto an Earth-like planet in the habitable zone of M-dwarf, K- and G-type stars, as a function of the spacing between scattering planets (in units of mutual Hill radius). Typical values for the terrestrial planets in the Solar System, and the tightly-packed TRAPPIST-1 planets are added for reference. \textbf{Right panel.} The corresponding analytical predictions for the minimum velocity impacts onto the planet. Rough velocity thresholds, based off values for HCN survival \cite{ToddOberg2020} are added for reference. In both panels, the grey shaded region is an unstable region ($\Delta < \Delta_c$), for which planets will have close encounters in less than $\sim 10^2$ orbits.}
    \label{fig:Tiss_vimp_hill_spacing}
\end{figure}
In order to analytically predict the minimum impact velocities, we consider the comet's Tisserand parameter, conserved during three-body interactions with the scattering planets. The maximum Tisserand parameter (equation~\ref{eq:maximum_tiss_param}), corresponding to comets scattered between neighbouring planets, is shown in figure~\ref{fig:Tiss_vimp_hill_spacing}(a). We see the maximum Tisserand parameter decreases with the planetary separation ($\Delta$), and that the effect of the planetary architecture is more significant around low-mass stars. The corresponding minimum impact velocity, given by equation~\ref{eq:minimum_imp_vel} is shown in figure~\ref{fig:Tiss_vimp_hill_spacing}(b), and increases strongly with $\Delta$ for low-mass stars. For low-mass stars (0.1\,$M_\mathrm{Sun}$), the range of spacings considered corresponds to scattering directly from the snow-line, all the way up to having 6 planets spaced equally between the snow-line and habitable zone.

The dependence of both the maximum Tisserand parameter and minimum impact velocity on $\Delta$ is seen to be stronger around low-mass stars. Since the planet-to-star mass ratio is larger around low-mass stars, the width of the resonant regions around these planets will also increase. Consequently, fixed separations in mutual Hill radius correspond to larger distances in physical-space around low-mass stars, and therefore lower (higher) Tisserand parameters (impact velocities).

For the case of direct scattering between the snow-line and habitable zone (i.e planetary separations of $\sim$\,70\,$\Delta$) the minimum impact velocities reach $\sim$\,20\,km/s around low-mass stars. This high velocity is attributed to the larger ratio of the snow-line to the habitable zone around low-mass stars, and the (roughly) $M_\ast^{-1/2}$ scaling of the orbital velocity of habitable planets. As velocities exceed 20\,km/s, HCN survival decreases exponentially from $\sim$\,1\%, severely limiting the efficiency of cometary delivery. Above 25\,km/s impact blowoff from the atmosphere will prevent any successful cometary delivery \cite{ToddOberg2020}.

\begin{figure}[t!]
    \centering
    \includegraphics[width=\textwidth]{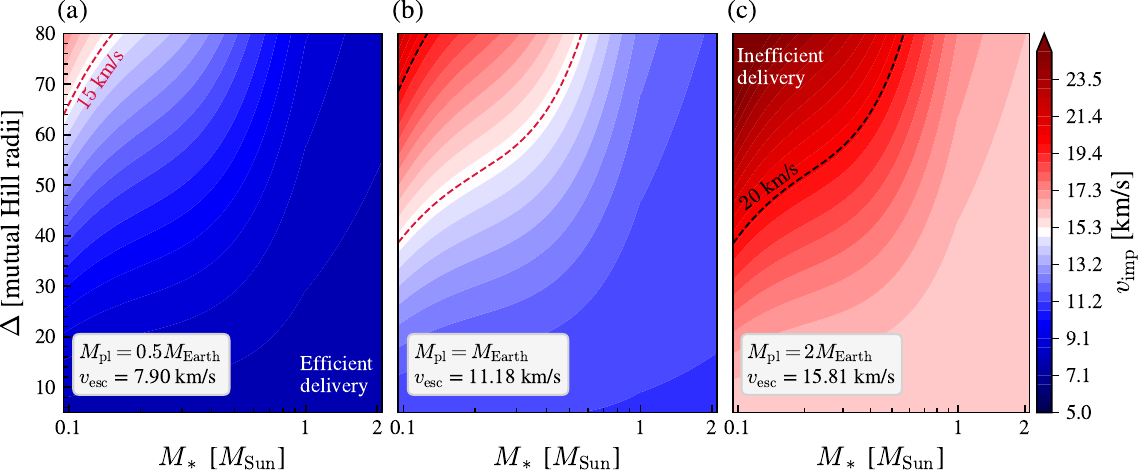}
    \caption{The minimum impact velocity onto an Earth-like planet in the habitable zone is calculated across a wide range of $\Delta$-$M_\ast$ parameter space. The planet's escape velocity is changed between the three panels. The survival of HCN is used as a proxy for the overall efficiency of cometary delivery. HCN delivery is most effective below 15\,km/s, decreasing sharply above 20\,km/s for which there will be very limited survival \cite{ToddOberg2020} Above 25\,km/s there will be no HCN survival due to impact blowoff from the atmosphere.}
    \label{fig:vimp_contours}
\end{figure}

In summary we find the effectiveness of cometary delivery will depend on the planet's escape velocity, the planetary architecture and the stellar-mass, as illustrated in figure~\ref{fig:vimp_contours}. Figure~\ref{fig:vimp_contours}(a) shows how reducing the planet's escape velocity is able to keep the impact velocity substantially below 15\,km/s, effectively erasing all effects of the planetary architecture and stellar-mass. For an Earth-like planet, the importance of the planetary architecture in the context of cometary delivery is almost completely dependent on the stellar-mass, as shown in figure~\ref{fig:vimp_contours}(b).

\subsection{Numerical results}
The results of our N-body simulations, as described in section~\ref{sec:num_sims}, are shown in figure~\ref{fig:KM_theory_simulation_comparison} for a range of stellar masses [0.1, 0.4, 1]\,$M_\mathrm{Sun}$. Each panel corresponds to a different stellar mass, illustrating the effects of the planetary architecture for three planetary spacings, $[10, 30, 50]\,R_{\rm H,m}$. The velocity distribution onto a habitable planet orbiting a 1\,$M_\mathrm{Sun}$ star is included in the background for comparison. As predicted, there is a large change in the minimum impact velocity as a function of $\Delta$, which our analytical model is able to accurately reproduce as indicated by the dashed lines.

There are a small number of impacts below our analytical predictions, which are a consequence of our assumption that comets must have planet-crossing orbits to be scattered by the planets in the system. Instead, there exists extended regions surrounding planets ($\sim$\,$R_\mathrm{Hill}$) within which particles can be scattered, increasing the available parameter space for the comets. We expect this effect to be more significant in systems with higher-mass planets and smaller planetary separations, as comets with low $v_\infty$ will have much longer interaction times with the planets, allowing impacts from much larger impact parameters\footnote{The conservation of angular momentum determines a maximum impact parameter for impacts, $b_\mathrm{max} = R_\mathrm{pl} v_\mathrm{imp} / v_\infty \sim R_\mathrm{pl} v_\mathrm{esc} / v_\infty$}. Some additional complexity arises as comets can be scattered backwards and forwards between just two planets, which can substantially alter their Tisserand parameter \cite{BonsorWyatt2012}. These are, however, minor effects, as seen in the low-velocity tails of figure~\ref{fig:KM_theory_simulation_comparison}, and are actually beneficial to the successful delivery of prebiotic molecules, so do not substantially affect the main conclusions we reach. 

We also see clear qualitative differences between the velocity distributions in figure~\ref{fig:KM_theory_simulation_comparison}, highlighting the fact that it is not just the minimum impact velocity that is affected by the planetary architecture and stellar-mass. A key takeaway is that there are a much larger fraction of very low-velocity impacts around Solar-mass stars, which suggests that comets would be able to deliver a significantly larger inventory of prebiotic molecules. The fraction of low-velocity impacts is only found to depend on the planetary architecture for planets around low-mass stars, as seen in figure~\ref{fig:KM_comparison_low_velocity_fracs}, which is supported by the general results of our analytical model.

An additional, striking feature of the velocity distributions, is that the spread in impact velocities is much larger for systems around low-mass stars. This is driven by the extended high-velocity tails, which remain almost unchanged by the planetary spacing. These are comets scattered onto highly eccentric orbits by the outer planets, which remain on such orbits before impacting the inner habitable planet. This has potentially significant implications for cometary delivery, as the survival rate decreases exponentially with impact velocity. Furthermore, the increased fraction of high-velocity impacts around low-mass stars may be prejudicial to the development, and persistence of life on these planets \cite{MaherStevenson1988}. We will discuss these changes to the velocity distributions, and the associated implications for cometary delivery in section~\ref{sec:discussion}.

\subsection{Summary of results}
\label{sec:summary_results}
\begin{figure}[t!]
    \centering
    \includegraphics[width=1.\textwidth]{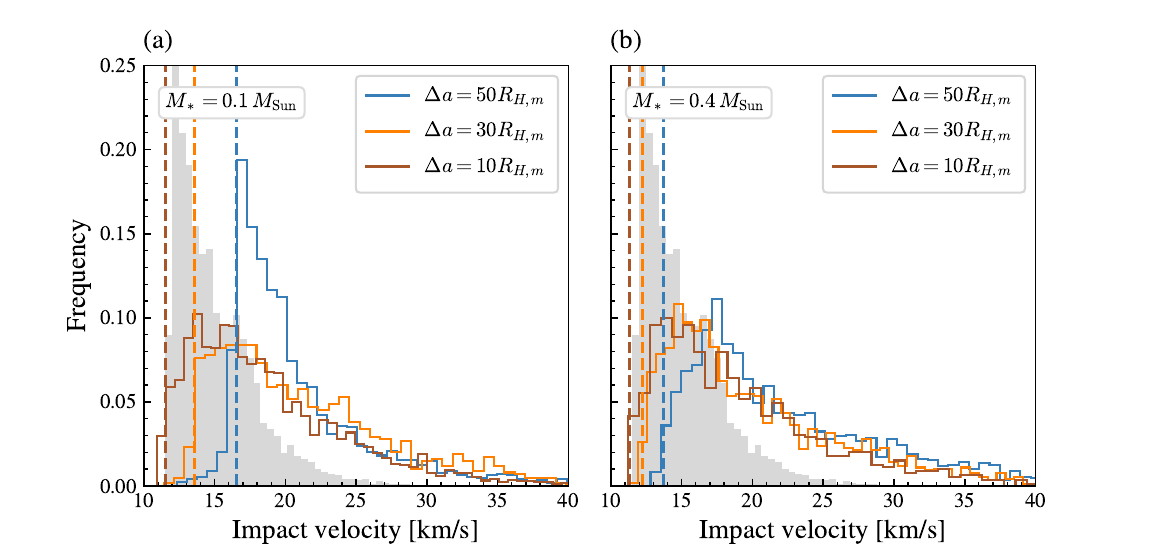}
    \caption{Impact velocity distributions from N-body test-particle simulations onto an Earth-like planet in the habitable zone of an 0.1\,$M_\mathrm{Sun}$ M-dwarf (a) and an 0.4\,$M_\mathrm{Sun}$ M-dwarf (b). The planetary spacing, in units of mutual Hill radius, is varied between $10\,R_{\rm H, m}$ and $50\,R_{\rm H, m}$ in each panel. This translates to between 2 and 8 planets spaced equally between the snow-line and habitable zone. The vertical dashed lines are the analytical predictions for the minimum velocity impacts (as seen in figure~\ref{fig:Tiss_vimp_hill_spacing}). For comparison, the velocity distribution for a planet around a G-type star (1\,$M_\mathrm{Sun}$, $\Delta a = 50 R_{H,m}$) is shaded grey in the background. The overall shapes of these distributions are very sensitive to the initial distribution of the test-particles, which we describe in section~\ref{sec:num_sims}.}
    \label{fig:KM_theory_simulation_comparison}
\end{figure}
We have demonstrated analytically and numerically that very low velocity impacts onto Earth-like planets are possible (impossible) around Solar-type (M-dwarf) stars for comets scattered directly from beyond the snow-line. We show, however, that the presence of a number of planets between the habitable zone and snow-line is able to significantly reduce the minimum impact velocity, independent of stellar-mass. The dependence of the minimum impact velocity on the planet's escape velocity, the stellar-mass and the planetary architecture is shown in figure~\ref{fig:vimp_contours}. The results of our N-body simulations, figures~\ref{fig:KM_theory_simulation_comparison} and~\ref{fig:KM_comparison_low_velocity_fracs}, demonstrate that the overall velocity distributions onto habitable planets are also very sensitive to the stellar-mass and planetary architecture. Importantly for the possibility of cometary delivery, there will be substantially more low-velocity impacts onto planets in tightly-packed planetary systems and around higher-mass stars.

\section{Discussion}
\label{sec:discussion}
\subsection{For which planets might cometary delivery be important?}
\label{sec:general_delivery_discussion}
As summarised in section~\ref{sec:summary_results}, we find the minimum impact velocity of comets onto habitable planets is reduced for i) planets with lower escape velocities (lower mass), ii) planets orbiting higher-mass stars, and iii) planets in tightly-packed systems, for which comets are delivered on low eccentricity orbits. The survival of prebiotically interesting molecules in comets depends on both the impactor's size and velocity and so, whilst remaining agnostic to the size-frequency distribution, we are able to make robust statements about the relative efficiency of cometary delivery. 

The most promising planets for any cometary delivery at low impact velocities will be found around F- and G-type stars, largely irrespective of the planetary architecture (see figure~\ref{fig:vimp_contours}). The simulations presented here indicate that the delivery of the prebiotically relevant molecule HCN, favoured for impacts below 15\,km/s (low impact angle, small radius) \cite{ToddOberg2020}, is possible, but not common for Earth mass planets in the habitable zone of Sun-like stars (see figure~\ref{fig:KM_theory_simulation_comparison}). 

Delivery at low velocities, including sufficiently gentle impacts for there to be the successful delivery of an appreciable amount of more complex prebiotic molecules, such as amino acids \cite{PierazzoChyba1999}, will always occur more frequently for lower mass planets due to their lower escape velocities, as seen in figure~\ref{fig:vimp_contours}. A consequence of this conclusion is that comets could have delivered a larger, and potentially more complex initial prebiotic inventory to Mars than Earth. The same is also true for planets around massive stars (i.e. $M_\ast > M_\mathrm{Sun}$), given the orbital velocity in the habitable zone decreases roughly as $M_\ast^{-1/2}$. The short main sequence lifetimes, and inhospitable ionising winds from A-, B- and O- type stars \cite{MaederMeynet1998, KudritzkiPuls2000} suggest that there are other reasons why life may not be favoured, pointing to F-type stars as the most promising high-mass stars to support life. A number of unique challenges for the emergence and evolution of life (post cometary delivery) remain for planets around F-type stars, including enhanced UV-fluxes \cite{Cockell1999}, and the migration of the habitable zone throughout the stellar main sequence \cite{Sato2014}.
\begin{figure}[t!]
    \centering
    \includegraphics[width=\textwidth]{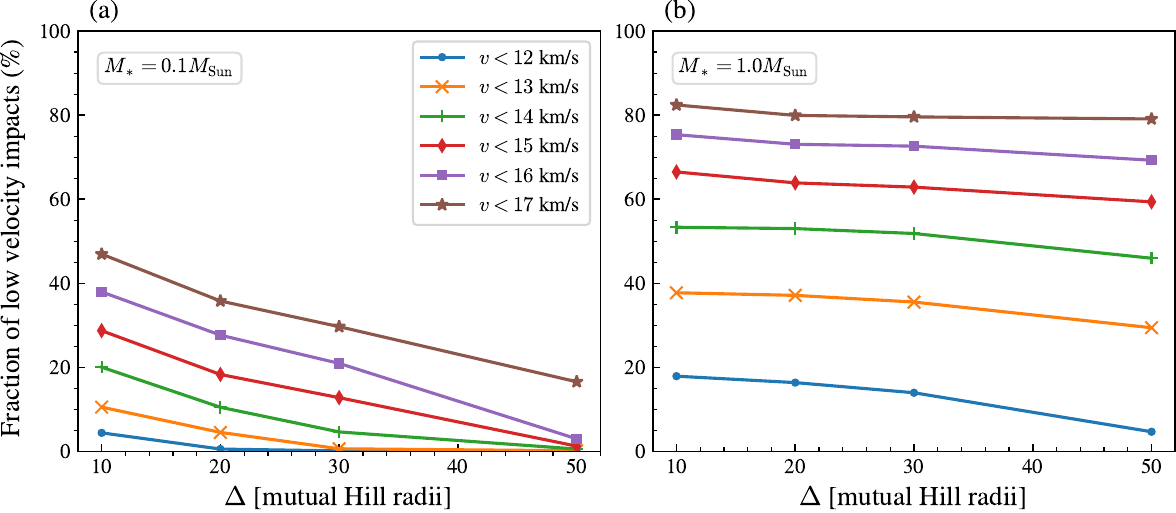}
    \caption{The fraction of low-velocity impacts as a function of the planetary spacing for a number of velocity thresholds are shown for a planet around both an M-dwarf (\textbf{left panel}) and G-type star (\textbf{right panel}). Both the minimum velocity and the full velocity distribution shift to lower velocities when the planetary spacing is decreased, seen clearly for planets around M-dwarfs. A similar trend is only seen for the very lowest velocity impacts around G-type stars, as predicted by our analytical model (see e.g. figure~\ref{fig:Tiss_vimp_hill_spacing}). For reference, using \cite[equation 6]{ToddOberg2020}, HCN survival is enhanced by roughly 100\% for impacts at 12\,km/s compared to 17\,km/s. These changes in the velocity distribution at small $\Delta$ therefore correspond to significant increases in HCN survival.}
    \label{fig:KM_comparison_low_velocity_fracs}
\end{figure}

Successful cometary delivery at low velocities is significantly more challenging for planets around low-mass stars. For comets scattered directly from the snow-line, minimum impact velocities can exceed 20\,km/s onto Earth-like planets. For context, impacts at 20\,km/s for a 6\,km comet result in approximately 0.2\,\% HCN survival, which decreases to 0\,\% above 25\,km/s, where significant atmospheric blowoff is expected \cite{ToddOberg2020}. We show, however, that this minimum impact velocity can be significantly reduced in multiple-planet systems, suggesting that successful cometary delivery is possible in tightly-packed systems, often observed around low-mass stars. The fraction of low velocity impactors increases in more tightly-packed systems (figure~\ref{fig:KM_comparison_low_velocity_fracs}), which further enhances the potential efficiency of cometary delivery in these scenarios. 

TRAPPIST-1 is an example of a tightly-packed planetary system around a low-mass M-dwarf (0.0898\,$M_{\rm Sun}$ \cite{Agol2021}), with 3 planets in the habitable zone (e, f, and g)  \cite{Gillon2017}. Proxima Centauri is another low-mass M-dwarf (0.12\,$M_{\rm Sun}$) also with a habitable zone planet, Proxima Centauri b \cite{AngladaEscude2016}. A crucial distinction is that Proxima Centauri b has only one known external companion, the candidate planet Proxima Cenaturi c, which is on a distant orbit beyond the snow-line \cite{Damasso2020}. In the case this interior region truly lacks any additional planets and debris, this resulting architecture would lead to significantly higher impact velocities when compared to the TRAPPIST-1 planets. This would, in turn, dramatically reduce the efficiency of any cometary delivery. We note that these results are in agreement with the general picture of fast litho-panspermia in the TRAPPIST-1 system presented in \cite{Krijt2017}. 

The dynamical timescales around low-mass stars are much shorter than Solar-type stars, whilst the pre-main sequence cooling phase is significantly longer around low-mass stars \cite{RamirezKaltenegger2014}, a prolonged period of enhanced stellar activity and high XUV-fluxes \cite{Hawley2014, Baraffe2015}. These effects combine such that the late-bombardment of habitable planets will end before the star reaches the main sequence \cite{Clement2022, Lichtenberg2022}. These planets will have likely exceeded the runaway greenhouse threshold \cite{Bolmont2017}, and so any cometary delivery during this period would occur during the planet's magma ocean epoch \cite{Lichtenberg2022}, unconducive to the initial emergence of life via the comet pond scenario. More promisingly, recent work indicates that volatile-rich debris is ubiquitously scattered into exo-asteroid belts during the planet formation process \cite{Clement2022}. This would support our general picture of scattering from a volatile-rich reservoir beyond the snow-line, and allow for delayed cometary delivery across a wide range of stellar-masses. 

\subsection{The effects of the overall velocity distribution}
\label{sec:overall_vel_dist}
Whilst crucial for cometary delivery, low-velocity impactors constitute only a small fraction of the overall population of impacting bodies (as seen in figure~\ref{fig:KM_theory_simulation_comparison}). The fraction of high-velocity impacts increases significantly around low-mass stars (see figure~\ref{fig:KM_comparison_low_velocity_fracs}), which remain even in tightly-packed systems. These impactors have potentially negative implications for the habitability of planets around low-mass stars, even in the case of successful cometary delivery, which we will now discuss.

Comet ponds are particularly exposed settings for abiogenesis, sensitive to both their immediate environment and also global climatic conditions, so are very sensitive to a planet's impact history \cite{MaherStevenson1988}. Increased fluxes of high-velocity impactors are therefore likely to be prejudicial to the subsequent emergence of life. High-velocity impacts are capable of shock heating atmospheres to very high temperatures ($\sim$\,100\textdegree\,C), which can be followed by prolonged periods of cooling to temperatures below 0\textdegree\,C due to the shielding effects of impact ejecta \cite{Toon1982}. These ejecta blankets can also seriously disrupt the chemical evolution of comet ponds due to, and not limited to, acidification, heating and mixing \cite{MaherStevenson1988}.

Moving beyond the sterilising potential of high-velocity impacts, modelling the overall velocity distribution is crucial for characterising a planet's atmospheric evolution. Atmospheric evolution will depend on the relative efficiency of volatile delivery to loss from atmospheric stripping, which will be dependent on impactors' velocity distribution, size distribution and composition (e.g. \cite{Schlichting2015}). Habitable planets around low-mass stars will therefore be more susceptible to atmospheric erosion (see also \cite{Wyatt2020, Sinclair2020}), which is largely driven by impacts above twice the planet's escape velocity \cite{MeloshVickery1989}. The sensitivity of atmospheric evolution calculations to the impactors' velocity distribution means that tailored N-body simulations are necessary to make accurate conclusions about individual planetary systems, as demonstrated in \cite{Kral2018}. 

High-velocity impactors might, however, support other origins scenarios that do not require the initial emergence of life from small surface-level ponds. High-velocity impacts can enrich a planet's volatile inventory through the shock synthesis of prebiotic feedstock molecules \cite{ChybaSagan1992}, in which a number of chemicals (including HCN) can be synthesised in the impactor's plasma. As with many other HCN formation pathways, this effect is found to be most effective in reduced atmospheres \cite{Ferus2017, Parkos2018}. For the early-Earth it is thought that at least one impactor will have been able to generate prebiotically relevant concentrations of HCN, even for an oxidised early-atmosphere \cite{Parkos2018}. The increased fraction of high-velocity impacts onto habitable planets around low-mass stars could therefore indicate more effective endogenous HCN production, a tentative conclusion contingent on the oxidation state of the planet's early-atmosphere. A separate study would therefore be required to understand the efficiency of impact-driven HCN production in different environments, and allow for a comparison with cometary delivery.

\subsection{Model limitations}
\label{sec:model_limitations}
To analytically explore how the stellar-mass and planetary architecture affects the impact velocity distribution onto habitable planets, we necessarily make a number of simplifying assumptions. This enables us to draw some broad conclusions about the importance of cometary delivery in a wide range of planetary environments. The main assumptions we make, and the potential implications for our conclusions, are summarised here.

In this work we have considered highly idealised planetary architectures, with equal mass planets spaced equally in terms of their mutual Hill radius. This may be reflected in the observational population of peas-in-a-pod systems \cite{Weiss2022}, but our architectures are significantly more idealised than these observed systems. Previous work has demonstrated, however, that a variety of architectures are able to scatter exocomets onto the inner planets \cite{Marino2018}, with the scattering efficiency dependent on both the planet masses and spacings. This suggests a robustness to minor changes in the planetary architecture, with the details of the impacting flux and velocity distribution onto habitable planets changing on a system-by-system basis. 

As discussed in Section~\ref{sec:planetary_architecture}, we do not specify the planetary architecture beyond the snow-line, however the presence of an external giant planet could be crucial in perturbing comets onto habitable zone-crossing orbits \cite{DencsRegaly2019}. Given the efficiency of pebble accretion beyond the snow-line \cite{Bitsch2019}, we argue this justifies our choice of architectures, as external giant planets would be able to provide an inwards cometary flux. The presence of a giant planet could however effectively prevent inwards scattering depending on its location in $(a_\mathrm{pl}, M_\mathrm{pl})$ space \cite{Wyatt2017}. We note here that the dynamics of every individual planetary system will be unique, with particular (in)efficiencies in scattering dependent on the exact orbital locations of planets, and that both resonant and secular effects could in principle produce results that differ significantly from the trends we present in this paper.

The results of our N-body simulations are very sensitive to the assumptions we make regarding the initial distribution of the test-particles. Our choice of a dynamically cold distribution of test-particles gives us an upper limit on the fraction of low-velocity impacts, providing an optimistic picture of the potential efficiency of cometary delivery. If, however, the outer cometary belt is dynamically excited, low velocity impacts will be unlikely. Our assumption of a planet at the inner edge of a static snow-line is also highly idealised, with the location of the snow-line evolving quickly during initial planet formation before moving slowly outwards at late times \cite{ZhangJin2015}. Every system will have a different architecture, but if instead a giant planet beyond the snow-line is the dominant source of volatile-rich comets, this would have the same effect as changing the initial Tisserand parameter distribution, shifting the velocity distribution towards higher velocities. We also acknowledge that in practice outer belts are likely to host very diverse populations of volatile-rich bodies, as seen in the solar system (e.g. \cite{Nesvorny2015}). These do not necessarily need to be scattered from the snow-line, as near-Earth asteroids have also been found to host a suite of prebiotically relevant molecules, with amino acids, amines and the RNA nucleobase Uracil found on the carbonaceous asteroid Ryugu \cite{Oba2023, Naraoka2023}.

Finally, the successful cometary delivery of prebiotic molecules is also contingent on a number of physical parameters, which include most significantly the comet's radius, the atmospheric density, and the location of impact \cite{Chyba1990, PierazzoChyba1999}. For example, the atmospheric density is crucial in determining the efficiency of aerobraking, which in principle is able to sufficiently decelerate large comets in very dense ($\sim$10\,bar) atmospheres and facilitate increased organic survival \cite{Chyba1990}. Smaller bodies (i.e <0.5\,km) will violently disrupt at high altitudes due to comets' low material strengths \cite{Chyba1993}, and so it is likely that in this size range these `airbursts' will prevent successful cometary delivery \cite{BosloughCrawford2008}. Similarly, organic survival will decrease for larger radius comets, and for impacts onto the continental crust, rather than deep ocean \cite{PierazzoChyba1999}. In light of the recent atmospheric retrievals of K2-18b, which is potentially a temperate, water-rich sub-Neptune \cite{Madhusudhan2023}, a study of cometary delivery onto `Hycean' planets may be of interest, given their deep atmospheres and oceanic surfaces. Cometary delivery may be particularly important, given that high pressure ices below the oceans would likely preclude access to volatile elements in the silicate core \cite{Madhusudhan2023_chemical_conditions}. In this work, however, we remain agnostic to these effects in our analytical model as they will change on a planet-by-planet basis, and require additional assumptions, such as the comets' size-frequency distribution. These effects will, however contribute additional scatter onto our conclusions regarding the importance of cometary delivery.

\subsection{Observational outlook}
\label{sec:biosignatures_discussion}
Exoplanets can be seen as a diverse array of laboratories, which we can use to study different origins-of-life scenarios (e.g. \cite{Rimmer2023}). In the coming decade, the HabEx-PLATO-LIFE epoch will facilitate the detection and atmospheric characterisation of a census of potentially habitable Earth-mass planets, extending even to planets around Sun-like stars. These are prime candidates for supporting Earth-like conditions, improving the prospects of detecting the first signs of life outside our home planet. This work suggests that in the scenario that the cometary delivery of prebiotically relevant molecules at low velocities is important for the origin of life, there will be a correlation between the presence of biosignatures and the following populations,
\begin{enumerate}
    \item \textbf{Exoplanets with low escape velocities.} A strict upper bound at around 25\,km/s would indicate a very strong reliance on cometary delivery.
    \item \textbf{Exoplanets in tightly-packed systems.} This should be a stronger effect for the population of planets around low-mass stars (see figure~\ref{fig:vimp_contours}).
    \item \textbf{Exoplanets around high-mass stars} ($M_\ast \gtrsim 1 M_\mathrm{Sun}$)\textbf{.} This effect would be seen roughly independent of planetary spacing.
\end{enumerate}
Alternatively, the absence of such a correlation would strongly indicate that cometary delivery at low velocities is unimportant for the origins of life on rocky exoplanets.

\section{Conclusions}
\label{sec:conclusions}
This paper studies the potential of cometary impacts to deliver the initial prebiotic inventory required for the origins of life on rocky exoplanets. We consider the scattering of comets by an idealised planetary system, and derive simple analytical expressions for the minimum impact velocity onto habitable planets. This allows us to draw a number of broad conclusions about the importance of cometary delivery across a wide range of planetary environments.

We find that minimum impact velocities are reduced for planets with lower escape velocities, planets in tightly-packed systems, and for planetary systems around high-mass stars. Furthermore, the results of our N-body simulations demonstrate that the overall velocity distribution of impactors onto habitable planets is very sensitive to both the stellar-mass and planetary architecture, with the fraction of low-velocity impacts increasing significantly for planets around Solar-mass stars, and in tightly-packed systems. It will be these populations of exoplanets where successful cometary delivery of prebiotic molecules is most likely to be successful, with significant implications for the resulting prebiotic inventories due to the exponential decrease in survivability with impact velocity.

Our results highlight the importance of understanding a planet's bulk properties, the stellar-mass and the surrounding planetary environment, as all of these factors individually can drastically affect a comet's minimum impact velocity. If future space missions, such as HabEx/LIFE, characterise the atmospheres of rocky planets around stars with a range of stellar masses, this work predicts that if cometary delivery is important for the origins of life, the presence of biosignatures will be correlated with increased stellar-mass, decreased planetary mass, and decreased planetary spacing (i.e. tightly-packed systems). The absence of such correlations would suggest that alternate pathways are required to produce the initial inventories of prebiotic feedstock molecules on rocky exoplanets.

\enlargethispage{20pt}

\ack{We thank Matthew Clement and an anonymous reviewer for insightful comments that have greatly improved this manuscript. RJA acknowledges the Science and Technology
Facilities Council (STFC) for a PhD studentship. AB acknowledges the support of a Royal Society University Research Fellowship,  URF\textbackslash R1\textbackslash 211421. Simulations in this paper made use of the REBOUND code which can be downloaded freely at \url{https://github.com/hannorein/rebound}.
\newline \textit{Software acknowledgements:} \textsc{Rebound} \cite{ReinLiu2012}, \textsc{Numpy} \cite{numpy2020}, \textsc{Matplotlib} \cite{matplotlib2007}, \textsc{Astropy} \cite{astropy2022}.}


\vskip2pc

\bibliographystyle{RS}
\bibliography{sample}

\end{document}